\documentclass[fleqn,usenatbib]{mnras}

\usepackage{newtxtext,newtxmath}

\usepackage[T1]{fontenc}
\usepackage{ae,aecompl}

\usepackage{graphicx}	
\usepackage{amsmath}	
\usepackage{amssymb}	

\title[Halo mass in axion cosmology]{Minimum star-forming halo mass in axion cosmology}

\author[J. M. Sullivan et al.]{
James M. Sullivan,$^{1,3}$\thanks{E-mail: jsull3@utexas.edu}
Shingo Hirano$^{1,2}$ and
Volker Bromm$^{1}$
\\
$^{1}$Department of Astronomy, University of Texas at Austin, Austin, TX 78712, USA\\
$^{2}$Department of Earth and Planetary Sciences, Faculty of Sciences, Kyushu University, Fukuoka, Fukuoka 819-0395, Japan\textbf{}\\
$^{3}$Department of Astronomy, University of California, Berkeley, CA 94720, USA\\
}

\date{Accepted XXX. Received YYY; in original form ZZZ}

\pubyear{2018}

\begin{document}
\label{firstpage}
\pagerange{\pageref{firstpage}--\pageref{lastpage}}
\maketitle

\begin{abstract}

Elucidating the particle physics nature of dark matter (DM) is one of the great challenges in modern science. The current lack of any direct DM detections in the laboratory heightens the need for astrophysical constraints, extending the search to DM models beyond the popular weakly interacting massive particle (WIMP) scenario. We here apply the classical Rees-Ostriker-Silk cooling criterion for galaxy formation to models with ultralight axion DM, also known as fuzzy dark matter (FDM). The resulting constraints provide a heuristic framework for upcoming observations, and our approximate analysis motivates the need for future, self-consistent simulations of FDM structure formation. We use observational constraints for the DM hosts of ultra faint dwarf (UFD) galaxies in the Local Group, together with the redshift constraints for the onset of primordial star formation from the recent EDGES 21-cm cosmology measurement, to illustrate this approach. We find that the existing constraints are straightforward to reconcile with standard $\Lambda$CDM, but disfavour FDM axion masses below $\sim 10^{-21}\,{\rm eV}/c^2$. The future potential for harnessing astrophysical probes of DM particle physics is compelling.
\end{abstract}



\section{Introduction}

The search for understanding the particle-physics nature of dark matter (DM) remains an extremely active topic in modern science.
The $\Lambda$ cold dark matter ($\Lambda$CDM) model successfully describes the observed large-scale structure of the Universe \citep[e.g.][]{Springel2005}.
On small scales, however, there are a number of discrepancies between the model prediction and observations \citep[see][and references therein]{2017ARA&A..55..343B}.
Moreover, the favoured DM particle candidate, the weakly interacting massive particle (WIMP), has yet to be detected, despite decade-long efforts for direct detection.
The confluence of these factors has led to a recent resurgence in attention to alternative DM models \citep[e.g.][]{Marsh2016}.

One alternative model, having attracted considerable attention recently, is fuzzy dark matter (FDM), a form of Bose-Einstein condensate DM that can be described by an ultra-light axion of mass $m_{\rm a} \gtrsim 10^{-22}\,{\rm eV}/c^2$ \citep{2000PhRvL..85.1158H}.  
FDM has the capacity to account for some of the small-scale problems associated with the $\Lambda$CDM model through quantum behaviour on galactic scales \citep[e.g.][]{2017PhRvD..95d3541H}, suppressing structure formation on small scales (wavenumbers $k \gtrsim 10 \ {\rm Mpc}^{-1}$).
Because the suppression scale sensitively depends on model parameters, in particular $m_{\rm a}$, it is crucial to assess whether the predicted minimum scale for structure formation is consistent with observations, such as the occurrence of dwarf galaxies in the Local Group.

An important concept in the theory of galaxy formation is the Rees-Ostriker-Silk (ROS) cooling criterion \citep{1977MNRAS.179..541R,1977ApJ...211..638S}. This condition identifies the minimum mass for DM haloes to be able to host star formation. When applied to the first structures \citep{Bromm2013}, within $\Lambda$CDM one arrives at `minihaloes' of mass $\sim 10^6\,{\rm M}_{\odot}$ and collapse redshifts of $z\sim 20-30$, where the first, so-called Population~III (Pop~III), stars form \citep[e.g.][]{Haiman1996,Tegmark1997}.
Given the small-scale sensitivity of DM models, the minimum threshold mass for structure formation offers a natural laboratory for DM particle physics. Specifically, we here evaluate the ROS criterion for FDM models, and compare with the observational constraints.
Since direct observation of Pop~III stars remain out of reach \citep[but see][]{Jaacks2018}, we turn to indirect signatures to provide constraints on the minimum halo mass. 

Ultra-faint dwarf (UFD) galaxies are the lowest-mass observable structures in the Universe today that can be used to infer properties about their host DM haloes. 
The lowest-mass UFDs, e.g. 
Segue~I \citep{Simon2011}, 
Segue~II \citep{2009MNRAS.397.1748B,2013ApJ...770...16K}, and Willman~I \citep{2011AJ....142..128W}, have mass-to-light ratios as high as $3000$, and are thus strongly dark matter dominated \citep[e.g.][]{Willman2012}.
As a result, their velocity dispersions are relatively high, and their masses are measured through kinematic studies of constituent stellar velocities.
Typically, UFDs have stellar masses of order $M_{*} \sim 10^{2}-10^{3}\,{\rm M}_{\odot}$, and are thought to have DM halo masses of order $M_{\rm halo} \sim 10^{7}-10^{9}\,{\rm M}_{\odot}$ \citep{2011ARA&A..49..373B}.
The lowest-mass UFDs then provide an upper limit for the minimum star-forming halo mass, if taking into account its subsequent mass accretion after star formation has ceased \citep[e.g.][]{Bovill2009,Milosavljevic2014}. 

Furthermore, we can place empirical constraints on halo formation, or virialization, times.
First, since UFDs also have the property of being galactic `fossils', they likely formed before reionization \citep[e.g.][]{2017ApJ...848...85J,Corlies2018}.
Second, another timing constraint can be obtained by probing the redshifted 21-cm signal from neutral hydrogen in the early Universe \citep{Furlanetto2006,Pritchard2012}. Recently, the Experiment to Detect the Global EoR Signature (EDGES) has discovered a global absorption trough in the 21-cm line \citep{2018Natur.555...67B}. This spectral feature implicates the early onset of Pop~III star formation, at redshifts $z\gtrsim 17$, which is natural within standard $\Lambda$CDM, but may challenge other DM models. In this
paper, we ascertain whether FDM scenarios are consistent with  observations,
considering the combined limits on halo mass and formation times.
This exploratory analysis is very timely with regard to the growing activity in UFD galaxy surveys, and pioneering 21-cm cosmological observations \citep[e.g.][]{Cohen2017,Barkana2018,Madau2018}.

\section{Physics of structure formation}
\label{phys_structure}

We briefly describe the basic physics of structure formation in the early Universe. Specifically, we apply the classical ROS cooling criterion to FDM models, focusing on the currently most promising mass range for ultralight axions, $m_{\rm a} = 10^{-22}$ and $10^{-21}\,{\rm eV}/c^2$. We will in turn discuss the collapse of DM haloes and the cooling properties of primordial gas when falling into them.

\subsection{Dark matter collapse}

Standard $\Lambda$CDM structure formation posits that DM haloes are seeded by overdense regions in the early Universe, where the overdensity exceeds the cosmic mean ($\delta > \bar{\delta}$). Such patches correspond to random Gaussian fluctuations in the initial cosmological density field, possibly arising from quantum noise in the inflationary fireball.
From linear theory, these overdensities are described in terms of the root-mean-square (rms) fluctuation on a given mass scale $\sigma(M)$, i.e. $\delta \sim \nu \sigma(M)$. Dense peaks eventually collapse and decouple from the expanding Universe to form virialized structures (DM haloes), once the overdensity exceeds a critical value of order unity. Pop~III stars typically emerge in $\lesssim 4\sigma$ peaks with halo masses $M_{\rm{vir}} \approx 10^{6}\,{\rm M}_{\odot}$, collapsing (or virializing) at $z_{\rm coll}\sim 20$. UFDs, on the other hand, form in somewhat more massive host haloes at $z_{\rm coll}\lesssim 15$, corresponding to $\lesssim 2\sigma$ peaks. 

In FDM cosmology, the primordial power-spectrum at small scales is truncated and the amplitude of small-scale fluctuations decreases \citep{Marsh2016}, resulting in the delayed formation of early, small-scale objects such as Pop III stars and UFDs.

\subsubsection{Quantum pressure}

As an additional requirement for collapse, 
halo masses must overcome the cosmological Jeans mass
\begin{equation}
M_{\rm J} = \frac{\pi}{6} \ \frac{c_{\rm s}^{3}}{G^{3/2} \rho^{1/2}},
\label{eq:jeans}
\end{equation}
where $c_{\rm s}$ is the gas sound speed in the intergalactic medium (IGM), $G$ Newton's constant, and $\rho$ the total mass density. In standard $\Lambda$CDM cosmology, this criterion is typically fulfilled for first structure formation in the cold, pre-reionization IGM, whenever the ROS cooling criterion is met (see Section~\ref{region}). FDM models, however, introduce a non-local quantum pressure in addition to the classical thermal gas pressure. The Jeans condition then needs to be modified accordingly.

The axion quantum pressure, due to the gradient of the density field, is \citep{2017MNRAS.471.4559M}
\begin{equation}
p_{\rm QM} = - \frac{\hbar^{2}}{4 m_{\rm a}^{2}} \rho_{\rm DM} \nabla \otimes \nabla \rm{ln} \rho_{\rm DM},
\end{equation} 
where $\hbar$ is the reduced Planck constant, and $\rho_{\rm DM}$ the DM mass density.
This pressure is repulsive in the early Universe and acts to counter the effect of gravity, possibly delaying the formation of the host DM haloes further, beyond what is already expected from small-scale power spectrum suppression. 
The quantum pressure directly acts only on the DM component, inducing a relative velocity between the DM and baryonic fluids in a given region.
We evaluate the effect of quantum pressure employing the methodology of \citet{2011ApJ...730L...1S}, used to explore the early structure formation delay due to the supersonic DM-baryon streaming motion imprinted at the recombination epoch \citep{2010PhRvD..82h3520T}.

We now assess the importance of quantum pressure in the collapse of the host haloes, where the first stars form. For simplicity, we assume an equation of state $p_{\rm QM} = \rho v_{\rm QM}^{2}$, and roughly estimate the pressure as $p_{\rm QM} \approx \hbar^{2}/(4 m_{\rm a}^{2}) \rho R_{\rm vir}^{-2}$, where $R_{\rm vir}$ is the virial radius of the DM halo \citep{2001PhR...349..125B}. The relative velocity due to the quantum pressure then approximately is $v_{\rm QM} \approx \hbar/(2 m_{\rm a}) R_{\rm vir}^{-1}$, to be compared to the sound speed in the cold IGM, $c_{\rm s}\sim 1$\,km\,s$^{-1}$, and the virial velocity in the host halo
\begin{equation}
v_{\rm vir} \approx \sqrt{\frac{G M_{\rm vir}}{R_{\rm vir}}}\mbox{\ .}
\end{equation}

To evaluate these quantities, we use the results from recent structure formation simulations in FDM cosmology \citep{2018MNRAS.473L...6H}: for $m_{\rm a} = 10^{-22}\,{\rm eV}/c^{2}$, collapse occurs with $M_{\rm vir} \simeq 4 \times 10^{9}\,{\rm M}_{\odot}$ and $R_{\rm vir}\simeq 5$\,kpc at $z_{\rm coll} \simeq 6.5$, whereas for $m_{\rm a} = 10^{-21}\,{\rm eV}/c^{2}$, the simulations predict collapse with $M_{\rm vir} \simeq 2 \times 10^{8}\,{\rm M}_{\odot}$ and $R_{\rm vir}\simeq 0.95$\,kpc at $z_{\rm coll} \simeq 13$. 
With these values, we estimate $v_{\rm QM} \lesssim 2$\,km\,s$^{-1}$ for both axion masses, comparable to the cold IGM sound speed, and an order of magnitude less than the respective virial velocities $v_{\rm vir}\lesssim 50$\,km\,s$^{-1}$.
We can thus conclude that the axion quantum pressure has no significant impact on the onset of the first star formation, and that the small-scale suppression in the FDM power spectrum is the dominant effect.
In the following, we will therefore focus on the latter.

However, this justification for neglecting the quantum pressure breaks down for the highest-sigma peaks, where virial masses approach the mass of the solitonic core. Again, within order of magnitude, we estimate that $v_{\rm QM}/v_{\rm vir}\propto (m_{\rm a}^2R_{\rm vir}M_{\rm vir})^{-1/2} \sim 1$ is reached at $\nu\sim 4$ for $m_{\rm a} = 10^{-21}\,{\rm eV}/c^{2}$, and $\nu\sim 3$ for $m_{\rm a} = 10^{-22}\,{\rm eV}/c^{2}$.

\subsection{Baryonic cooling physics}

In order to undergo gravitational runaway collapse, the infalling gas has to dissipate its compressional energy via radiative cooling.
A convenient way to capture this highly-complex process is the classical
ROS criterion \citep{1977MNRAS.179..541R,1977ApJ...211..638S}, requiring that for runaway collapse to occur the free-fall time-scale,
$t_{\rm ff} = \sqrt{3 \pi/32 G \rho}$,
must be longer than the gas cooling time-scale,
$t_{\rm cool} = (3/2) n k_{\rm B} T/\Lambda$,
where $n$ is the gas number density, $T$ the gas temperature, and $\Lambda$ the volumetric cooling function, such that $t_{\rm cool} < t_{\rm ff}$. When evaluated, this criterion can be translated into the properties of a successfully collapsing DM halo, in particular its virial mass and redshift. The ROS criterion has been thoroughly investigated for first star formation, where cooling relies on molecular hydrogen \citep[][and references therein]{Bromm2013}. Since the thermodynamic behaviour in FDM models converges to the standard $\Lambda$CDM minihalo case at densities $n\gtrsim 10^2$\,cm$^{-3}$ (see fig.~3 in \citealt{2018MNRAS.473L...6H}), we here assume for simplicity that the evaluation of the ROS criterion leads to similar constraints for the Pop~III host halo independent of cosmology. This assumption is plausible to first order, but needs to be further validated in future fully-consistent cosmological simulation work.

\subsection{Region of first star formation}
\label{region}

\begin{figure}
	\includegraphics[width=\columnwidth]{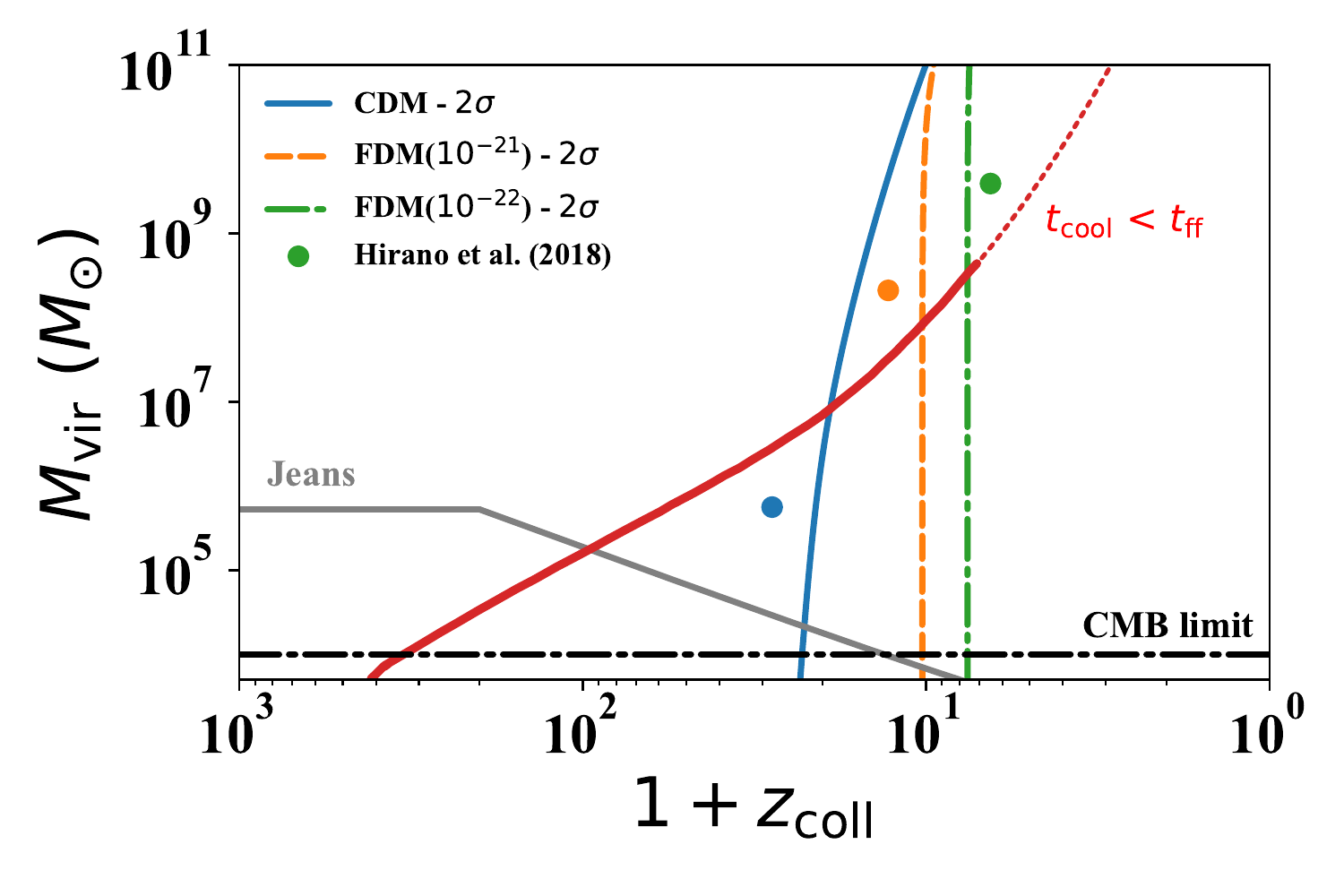}
    \vspace{-1cm}
    \caption{
Criteria for triggering primordial star formation, expressed in terms of halo virial mass vs. collapse redshift. Any successful host needs to have a mass exceeding the cosmological Jeans mass ({\it grey line}) and that required to fulfill the Rees-Ostriker-Silk criterion ({\it red line}). Towards lower redshift, the underlying modeling becomes increasingly uncertain ({\it dotted line style}). 
We also indicate the robust lower limit for the halo mass provided by the CMB temperature ({\it dot-dashed line}). The virial masses of DM haloes formed from $2 \sigma$ peaks in the $\Lambda$CDM universe ({\it blue}), as well as in FDM models with $m_{\rm a} = 10^{-21}\,{\rm eV}/c^2$ ({\it orange}) and $m_{\rm a} = 10^{-22}\,{\rm eV}/c^2$ ({\it green}) are plotted as a function of the collapse redshift. Primordial gas in DM haloes with masses above the minimum required by the two criteria may cool and collapse. The three dots represent select results found by cosmological simulations within the corresponding DM models \citep{2018MNRAS.473L...6H}.
    }
    \label{fig:ROS}
\end{figure}

In Fig.~\ref{fig:ROS}, we present the two requirements for successful runaway collapse of primordial gas, the cosmological Jeans mass and the ROS cooling criterion, in terms of virial mass of the DM halo, $M_{\rm vir}$, as a function of collapse redshift, $z_{\rm coll}$. For completeness, we also show a robust lower limit for the halo mass, where collapse is opposed by the thermal pressure of gas with a temperature coupled to the cosmic microwave background (CMB). As can be seen, however, the CMB limit is never important. Haloes with masses above those curves can in principle host the formation of the first stars, but the question now is whether such haloes are present at early times. Therefore, we next consider the Gaussian statistics of the growing density perturbations.

The three nearly vertical lines represent halo masses corresponding to $2\sigma$ peaks for CDM and two FDM models with $m_{\rm a} = 10^{-22}$ and $10^{-21}\,{\rm eV}/c^{2}$. These curves are calculated using the CDM power spectrum, utilizing the CAMB initial conditions generator \citep{Lewis:2013hha}, with an axion mass-dependant suppression given by equations~8 and 9 of \citet{2000PhRvL..85.1158H}. The normalized power spectra are then used to identify the parameters for the $2\sigma$ peaks.
Since the curve representing the ROS cooling criterion exhibits significant positive slope with decreasing redshift, the allowed region for primordial star formation in mass-redshift space is reduced for the FDM models, compared with the $\Lambda$CDM cosmology; thus, for FDM, higher halo masses are required that collapse later in time.
The canonical DM minihalo hosts for first star formation in the $\Lambda$CDM model, with typical masses of $\sim 10^{6}\,{\rm M}_{\odot}$, would fall below the ROS cooling curve in FDM cosmologies, and would thus not be able to trigger dissipative collapse. For comparison, we also present exploratory results found by recent cosmological simulations within different DM models \citep[coloured dots;][]{2018MNRAS.473L...6H}. These simulation results lie near the intersection of the ROS cooling line and the curves for the 2$\sigma$ peaks in the corresponding model. These are idealized results, as the simulations did not take quantum pressure into account. More complete simulations including quantum pressure have been performed \citep{2018arXiv180102320L,2018arXiv180108144N}, though they did not focus on the first stars. We emphasize again that the argument here aims to provide the big-picture theoretical context to guide upcoming observations. More precise predictions have to be derived with future simulations, where the physics of runaway collapse in FDM cosmologies is worked out in a self-consistent fashion.

\begin{figure}
	\includegraphics[width=\columnwidth]{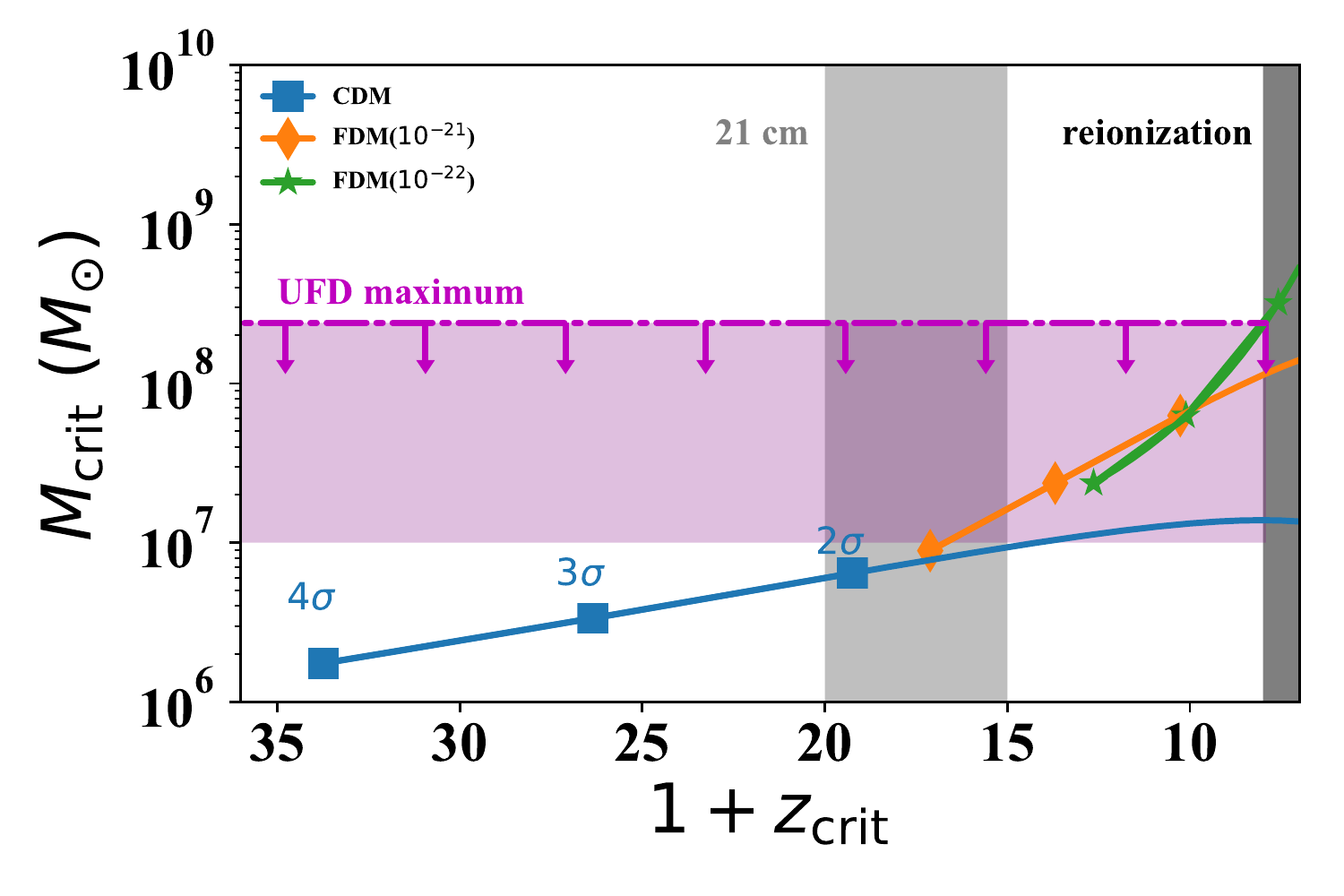}
    \vspace{-1cm}
    \caption{
Confronting structure formation theory with observations. Shown is the critical (minimum) DM halo mass required for the runaway collapse of primordial gas vs. redshift for different cosmological models (see Fig.~\ref{fig:ROS}). The three coloured lines represent the critical halo properties for CDM ({\it blue}), as well as FDM models with $m_{\rm a} = 10^{-21}\,{\rm eV}/c^2$ ({\it orange}) and $m_{\rm a} = 10^{-22}\,{\rm eV}/c^2$ ({\it green}).
The symbols indicate $2 \sigma$, $3 \sigma$, and $4 \sigma$ peaks, respectively. The magenta region summarizes estimates for the DM halo mass, hosting UFD galaxies in the Local Group. Complementary to this, the light-grey region delineates the redshift range inferred from the observed EDGES spectral trough in the 21-cm absorption profile \citep{2018Natur.555...67B}. The dark-grey region shows the epoch of reionization. As is evident, the combined empirical constraints disfavour FDM models with the lightest axion masses.
	}
    \label{fig:Obs}
\end{figure}

\section{Empirical constraints}
\label{obsv}

The crossing points of the ROS criterion and the halo mass formed from a given density fluctuation, given in Fig.~\ref{fig:ROS}, indicate the minimum star-forming halo mass, $M_{\rm crit}$, and the redshift when first star-formation occurs, $z_{\rm crit}$. In
Fig.~\ref{fig:Obs}, we summarize the critical halo mass vs. redshift behaviour for the cosmological models considered here, parameterized by the overdensity, $\nu \sigma$. To test the viability of the models, we confront these theoretical constraints with observations that are sensitive to the small-scale (low-mass) end of the power spectrum. Specifically, we consider dwarf galaxies in the Local Group, providing a fossil record of early star formation, and the recently discovered signature in the global spectrum of 21-cm radiation, constituting an in-situ, high-redshift probe of the first stars.

\subsection{Local Group dwarf galaxies}

The first constraint is the mass limit for the first star-forming haloes, $M_{\rm crit}$, such that at a given redshift, runaway cooling can only occur in sufficiently massive hosts. Observations of ultra-faint dwarf (UFD) galaxies provide non-trivial tests for any model of structure formation, as the model must allow star formation in low-mass host haloes, sufficiently early in cosmic history. Possibly, masses of the UFD progenitors may be as small as $M_{\rm vir} \approx 10^{7}\,{\rm M}_\odot$ at a redshift of $z \approx 10$ \citep{2015ApJ...807..154B}. However, observations of individual UFDs in the present-day Universe give relatively uncertain virial masses of order $M_{\rm vir} \approx 10^{8} - 10^{9}\,{\rm M}_\odot$ \citep[e.g.][]{Simon2011,2010MNRAS.406.1220W,2012MNRAS.422.1203B}, and a recent statistical analysis of local UFDs provides a conservative upper limit of the minimum UFD-hosting halo mass of $M_{\rm vir} \approx 2.5 \times 10^{8}\,{\rm M}_\odot$ \citep{2018MNRAS.473.2060J}.

We illustrate the plausible range of UFD halo masses in Fig.~\ref{fig:Obs}. Structure formation models represented by curves 
that lie beneath the upper limit of this region are consistent with the Local Group constraints. In the $\Lambda$CDM cosmology, $M_{\rm crit}$ is always below the region constrained by the UFD observations, regardless of overdensity $\nu \sigma$. This is in accordance with the standard picture of hierarchical structure formation. For the FDM model with the lighter axion, $m_{\rm a} = 10^{-22}\,{\rm eV}/c^{2}$, on the other hand, $M_{\rm crit}$ lies above the allowed UFD region for $2 \sigma$ peaks. In this case, the most frequently occurring haloes would form with masses larger than what is inferred for the UFD hosts. The resulting tension implies that existing UFD observations disfavour the lighter-axion FDM model.
Conversely, this tension does not exist in the model with heavier-axion FDM ($m_{\rm a} = 10^{-21}\,{\rm eV}/c^{2}$). Here, values of
$M_{\rm crit}$ for reasonable, not too unlikely, $\sigma$-peaks lie within or below the allowed UFD region.
Though it remains unclear if UFDs alone can rule out FDM entirely, further restrictions for the allowed models are anticipated through observations of lower-mass UFDs with upcoming large-scale surveys, carried out, e.g. with the Large Synoptic Survey Telescope \citep[LSST;][]{2009arXiv0912.0201L}.

\subsection{21-cm cosmology}

The second class of constraints arises from considering the formation redshift, $z_{\rm crit}$, of a given star forming system. The most direct, and stringent, constraint for the onset of star formation in the early Universe comes from the recent EDGES detection of the global 21-cm absorption trough \citep{2018Natur.555...67B}. They report that the 21-cm spin-flip transition in neutral hydrogen occurred at $z_{\rm EDGES} \approx 17.5$, which corresponds to the centre of the 21-cm absorption profile. The 21-cm signal requires the strong coupling of the spin temperature to the temperature of the cold IGM, mediated by Lyman-$\alpha$ radiation from the first stars through the Wouthuysen-Field effect \citep{Furlanetto2006}. Consequently, the EDGES measurement provides a lower redshift limit for the emergence of the first stars.

According to our analysis (Fig.~\ref{fig:Obs}), the $\Lambda$CDM model is fully consistent with the requirements from the 21-cm observations. All density peaks considered here form stars earlier than the central redshift of the EDGES trough. On the other hand, FDM models are challenged to accommodate the 21-cm data, for both axion masses explored in this paper. In the lighter-axion case, even the $4 \sigma$-peak halo reaches its collapse condition only after the EDGES absorption trough subsides, implying the impossibility to trigger the Wouthuysen-Field coupling sufficiently early. The heavier-axion case shows a similar result, but is somewhat less disfavoured than the lighter-axion one. Very rare $4 \sigma$-peaks may form at $z\approx z_{\rm EDGES}$, but no star formation is predicted before that. The redward ridge of the 21-cm absorption trough would require such earlier Pop~III stellar emission, again seriously challenging the model.

Recent work supports our findings disfavouring FDM models in the mass range $m_{\rm a} \approx 10^{-22}-10^{-21}\,{\rm eV}/c^{2}$, especially in light of the EDGES detection. 
By considering the Wouthuysen-Field coupling effect, \citet{2018arXiv180501253L} calculated an upper limit for the axion mass of $m_{\rm a} \gtrsim 5 \times 10^{-21}\,{\rm eV}/c^{2}$. Similarly, \citet{2018arXiv180500021S} found a slightly stronger axion mass constraint of $m_{\rm a} \approx 8 \times 10^{-21}\,{\rm eV}/c^{2}$, using simulations of high-redshift galaxies in a neutral gas medium. 
Additionally, a somewhat lighter constraint of $m_{\rm a} > 1.5 \times 10^{-22}\,{\rm eV}/c^{2}$ has been placed on the axion mass through dynamical modelling of stellar stream thickening \citep{AmoriscoLoeb}.
Results from upcoming instruments such as the Hydrogen Epoch of Reionization Array \citep[HERA;][]{2017PASP..129d5001D} will provide further constraints from the cosmic dawn 21-cm absorption profile, thus further strengthening limits on the allowed FDM axion mass. 

Confirming the EDGES redshift constraint in future work is crucial for testing FDM models, as well as other scenarios that suppress small-scale structure, such as warm dark matter (WDM) models \citep[e.g.][]{Dayal2017}. If the future accepted redshift for triggering effective Wouthuysen-Field coupling were much lower than reported by EDGES, around $z \sim 10$, heavier-axion FDM models would still be tenable. 
The lighter-axion model would still be disfavoured in this case, since the $2 \sigma$-peaks are in the exclusion region provided by reionization, as the first stars must necessarily have formed by the reionization redshift. If the EDGES results were to hold, however, a large sector of FDM models would firmly be excluded.

\section{Summary and Conclusions}
\label{conclude}

By considering the first stars we are able to connect the physical nature of DM to current frontiers in astronomical observation. The current non-detection of WIMPs in particle detector experiments accentuates the urgent need for such astrophysical probes. This approach promises to be increasingly potent with the imminent arrival of the next generation of observational facilities, such as the {\it James Webb Space Telescope (JWST)} and the suite of extremely large, $30-40$\,m-class telescopes on the ground. To fully harness this program, further developing the basic theoretical framework of cosmological structure formation in non-WIMP scenarios is essential, filling in the missing pieces in the exploratory, heuristic picture presented here.

Intriguingly, the recent constraints from 21-cm cosmology, indicating the early onset of significant star formation activity, go against the grain of the extensive observational hints for the effective suppression of small-scale structure \citep[e.g.][]{2017ARA&A..55..343B}. Again, confirming the EDGES results is thus of crucial importance. Together with the fossil record provided by near-field cosmology, astrophysical probes will increasingly place stringent constraints on the landscape of particle physics. We may well witness another case in the long history of science, where astronomy drives the progress of physics at its most fundamental level.

\section*{Acknowledgements}
This work was financially supported by Grant-in-Aid for JSPS Overseas
Research Fellowships and JSPS KAKENHI Grant 18J01296 (SH), as well as National Science Foundation (NSF) grant AST-1413501 (VB).




\bibliographystyle{mnras}
\bibliography{references.bib}

\bsp	
\label{lastpage}
\end{document}